\documentclass[iop,apj,tighten]{emulateapj} 
\usepackage{apjfonts}
\usepackage{lineno} 
\usepackage{natbib}
\usepackage{epsf}
\usepackage{url}
\usepackage{color}
\usepackage{mathrsfs, amsmath}

\usepackage[plainpages=false, colorlinks=true, anchorcolor=blue, linkcolor=blue, citecolor=blue, bookmarks=false]{hyperref}
\newcommand{\Msun}{\ifmmode\mbox{M}_{\odot}\else$\mbox{M}_{\odot}$\fi}
\newcommand{\Rsun}{\ifmmode\mbox{R}_{\odot}\else$\mbox{R}_{\odot}$\fi}
\newcommand{\Mearth}{\ifmmode\mbox{M}_{\oplus}\else$\mbox{M}_{\oplus}$\fi}
\newcommand{\Rearth}{\ifmmode\mbox{R}_{\oplus}\else$\mbox{R}_{\oplus}$\fi}
\newcommand{\msR}{\mathscr{R}}

\newcommand{\msG}{\mathcal{G}}
\newcommand{\msF}{\mathcal{F}}

\def\be{\begin{equation}}
\def\ee{\end{equation}}
\newcommand{\bb}{\begin{bmatrix}}
\newcommand{\eb}{\end{bmatrix}}

\shorttitle{21 years timing of PSR J1713+0747.}
\shortauthors{Zhu et al.}

\begin{document}
\title{Testing Theories of Gravitation Using 21-Year Timing of Pulsar Binary J1713+0747}

\author{
W. W. Zhu\altaffilmark{1,2},
I. H. Stairs\altaffilmark{1},
P. B. Demorest\altaffilmark{3},
D. J. Nice\altaffilmark{4},
J. A. Ellis\altaffilmark{5},
S. M. Ransom\altaffilmark{6},
Z. Arzoumanian\altaffilmark{7,8},
K. Crowter\altaffilmark{1},
T. Dolch\altaffilmark{9},
R. D. Ferdman\altaffilmark{10}, 
E. Fonseca\altaffilmark{1}, 
M. E. Gonzalez\altaffilmark{1,11}, 
G. Jones\altaffilmark{12}, 
M. L. Jones\altaffilmark{13}
M. T. Lam\altaffilmark{9}, 
L. Levin\altaffilmark{13,14}, 
M. A. McLaughlin\altaffilmark{13}, 
T. Pennucci\altaffilmark{15}, 
K. Stovall\altaffilmark{16}, 
J. Swiggum\altaffilmark{13}
}
\altaffiltext{1}{\footnotesize Department of Physics and Astronomy,
6224 Agricultural Road, University of British Columbia, Vancouver, BC, V6T 1Z1, Canada;
zhuww@mpifr-bonn.mpg.de, istairs@phas.ubc.ca}
\altaffiltext{2}{\footnotesize Max-Planck-Institut f\"{u}r Radioastronomie, Auf
dem H\"{u}gel 69, D-53121, Bonn, Germany}
\altaffiltext{3}{National Radio Astronomy Observatory, P.~O.~Box 0, Socorro,
NM, 87801, USA}
\altaffiltext{4}{\footnotesize Department of Physics, Lafayette College, Easton, PA 18042, USA}
\altaffiltext{5}{\footnotesize Jet Propulsion Laboratory, California Institute of Technology, 4800 Oak Grove Dr. Pasadena CA, 91109, USA}
\altaffiltext{6}{\footnotesize National Radio Astronomy Observatory, Charlottesville, VA 22903, USA}
\altaffiltext{7}{\footnotesize Center for Research and Exploration in Space Science and
Technology and X-Ray Astrophysics Laboratory, NASA Goddard Space Flight
Center, Code 662, Greenbelt, MD 20771, USA}
\altaffiltext{8}{\footnotesize Universities Space Research Association, Columbia, MD 21046, USA}
\altaffiltext{9}{\footnotesize Department of Astronomy, Cornell University, Ithaca, NY
14853, USA}
\altaffiltext{10}{\footnotesize Department of Physics, McGill University, Montreal, QC H3A
2T8, Canada}
\altaffiltext{11}{Department of Nuclear Medicine, Vancouver Coastal Health
Authority, Vancouver, BC V5Z 1M9, Canada}
\altaffiltext{12}{\footnotesize Department of Physics, Columbia University, 550
W. 120th St. New York, NY 10027, USA}
\altaffiltext{13}{\footnotesize Department of Physics and Astronomy, West Virginia University, P.O. Box
6315, Morgantown, WV 26505, USA}
\altaffiltext{14}{Jodrell Bank Centre for Astrophysics, School of Physics and
Astronomy, The University of Manchester, Manchester M13 9PL, UK}
\altaffiltext{15}{Department of Astronomy, University of Virginia, P.O. Box
400325 Charlottesville, VA 22904-4325, USA}
\altaffiltext{16}{\footnotesize Department of Physics and Astronomy, University of New
Mexico, Albuquerque, NM, 87131, USA}




\begin{abstract}
We report 21-year timing of one of the most precise pulsars: PSR~J1713+0747. 
Its pulse times of arrival are well
modeled by a comprehensive pulsar binary model including its three-dimensional
orbit and a noise model that incorporates short- and
long-timescale correlated noise such as jitter and red noise. Its timing residuals
have weighted root mean square $\sim92$~ns. 
The new data set allows us to update and improve previous measurements
of the system properties, including the masses of the neutron star
($1.31\pm0.11$~\Msun) and the companion white dwarf ($0.286\pm0.012$~\Msun) as well as their parallax 
distance $1.15\pm0.03$~kpc.  
We measured the intrinsic change in orbital period, $\dot{P}^{\rm Int}_{\rm b}$, is $-0.20\pm0.17$~ps~s$^{-1}$,
which is not distinguishable from zero.
This result, combined with the measured $\dot{P}^{\rm Int}_{\rm b}$ of other
pulsars, can place a generic limit on potential changes in the gravitational
constant $G$.
We found that $\dot{G}/G$ is consistent with zero
[$(-0.6\pm1.1)\times10^{-12}$~yr$^{-1}$, 95\% confidence] and changes at 
least a factor of $31$ (99.7\% confidence) more slowly than the average expansion 
rate of the Universe. This is the best $\dot{G}/G$ limit from pulsar binary
systems.
The $\dot{P}^{\rm Int}_{\rm b}$ of pulsar binaries can also place limits on the putative coupling
constant for dipole gravitational radiation
$\kappa_D=(-0.9\pm3.3)\times10^{-4}$ (95\% confidence).
Finally, the nearly circular orbit of this pulsar binary allows us to constrain statistically
the strong-field post-Newtonian parameters $\Delta$, which describes the
violation of strong equivalence principle, and $\hat{\alpha}_3$, which
describes a breaking of both Lorentz invariance in gravitation and
conservation of momentum. We found, at 95\% confidence, $\Delta<0.01$ and
$\hat{\alpha}_3<2\times10^{-20}$ based on PSR~J1713+0747.
\end{abstract}

\section{Introduction}
\label{sec:intro}
We present 21-year timing of the millisecond pulsar (MSP) J1713+0747, which
was discovered in 1993 \citep{fwc93}. It is one of the brightest pulsars timed by the
North American Nanohertz Observatory for Gravitational Waves (NANOGrav;
\citealt{mcl13, dfg+13}), and has the smallest timing residual of all NANOGrav
pulsars \citep{dfg+13}.
The timing analysis reported in this paper will be incorporated into future pulsar timing array
projects.
Timing observations of this pulsar were reported previously in
\citealt{cfw94}, \citealt{lb01}, \citealt{vb03}, \citealt{sns+05},
\citealt{hbo06} and \citealt{vbc+09}.
We report new results that arise from a significant extension of the timing
baseline and use of wide-bandwidth, high resolution instruments that allow us
to model and account for pulse time variations due to dispersion in the interstellar medium (ISM)
(Section \ref{sec:dmx}) and small distortions of the pulsar's magnetosphere (Section \ref{sec:FD}). 

MSPs are very stable rotators due to their enormous
angular momentum. PSR~J1713+0747 is an MSP residing in a wide binary orbit with a white dwarf companion (Section \ref{sec:model}). 
The pulse arrival times of the pulsar are well fit by a binary model with
a nearly circular orbit. The masses of
the binary components can be inferred through the measurement of the mass
function and Shapiro delay \citep{sns+05}. 
The system's distance is well-measured through a timing parallax. We detect
a changing projected orbital semi-major axis due to the orbit's proper motion
on the sky. Through the rate of change of projected semi-major axis, we can infer 
the orientation of the orbit in the sky.
This is one of the few binaries in which the 3D-orientation of the
binary orbit can be completely solved. Using 21 years of data, we
refine the previously published measurements of these orbital parameters.
We observed apparent variation of the binary orbital period due to the Shklovskii effect 
and Galactic differential acceleration.
We also find a stringent constraint on the intrinsic variation of the orbital
period, which enable us to test alternative theories of gravitation.

The stability and long orbital period of the PSR~J1713+0747 binary make it an
excellent laboratory for observing the time variation of Newton's gravitational ``constant'' $G$. 
This interesting conjecture of $G$ varying on a cosmological timescale was first 
raised by \citet{dir37} based on his large-number hypothesis. 
Later this become a prediction of some alternative theories of gravitation. 
For example, the scalar-tensor theory \citep{jor55,jor59,fie56,bd61} 
modifies Einstein's equation of gravitation by coupling mass with
long-range scalar-tensor fields, and predicts that, as the universe expands,
the scalar field will also change, causing the effective gravitational constant to vary  
on the cosmological timescale. 
Similar ideas were also revisited by gravitational theories involving extra dimensions
\citep{mar84,ww86a}.
PSR~J1713+0747 is likely the best pulsar binary for testing the constancy of
$G$ thanks to its high timing precision and long orbital period. Using timing
results reported in this paper, we found a stringent generic upper limit on 
$\dot{G}$ (see Section \ref{sec:Gdot} for details). 

The PSR~J1713+0747 binary is also an excellent laboratory for testing 
the strong equivalence principle (SEP) and the preferred
frame effect (PFE; in this work, we constrain the putative post-Newtonian
parameter $\hat{\alpha}_3$ that characterizes a specific type of PFE; for
details see Section \ref{sec:sep}). 
The violation of SEP or the existence of non-zero $\hat{\alpha}_3$
could lead to potentially observable effects which cannot be accounted in the
context of GR, such as forced-polarization of the binary orbit. Some of these
effects are discussed in \cite{fkw12} and \cite{will14}. We can put stringent generic
constraints on alternative theories by observing
low-eccentricity pulsar binary systems. Section \ref{sec:sep} presents the
constraint on the violation of SEP and the significance of $\hat{\alpha}_3$ from our 21 years of observation of PSR~J1713+0747.

\section{Observations}
\subsection{Data acquisition systems}

Our data set consists of pulse timing observations of PSR J1713+0747 at the
Arecibo Observatory, from 1992 through 2013, and at the Robert C. Byrd Green Bank
Telescope (GBT), from 2006 through 2013, using several generations of
data acquisition systems as described below (Table \ref{tab:obs}).  The first 12 yr of
these data (1992 August through 2004 May) were previously reported by
\citet{sns+05} using the Mark III, Mark IV, and Arecibo-Berkeley Pulsar
Processor (ABPP) systems.  In
this work, we incorporate an extension of the Mark IV data set (reduced
using the same process as \citealt{sns+05}); data collected at Arecibo
from two newer systems; and data collected at the GBT.

The earliest observations (Mark III) were made using a single receiver operating over 
a limited bandwidth and hence intrinsic pulsar behavior cannot be separated from
effects of the ISM.  Subsequent observations at both
telescopes were made using two receivers at widely spaced frequencies
(1410 and 2380 MHz at Arecibo; 800 and 1410~MHz at the GBT) to allow for
separation of these effects.



\begin{deluxetable*}{lccccccc}

\tabletypesize{\footnotesize}
\tablewidth{0pt}
\tablecaption{\label{tab:obs} 21 year J1713+0747 observations. }
\tablehead{ \colhead{System} & \colhead{Alias\tablenotemark{*}} & \colhead{Observatory} & \colhead{Dates}  &\colhead{Number of}
&\colhead{Epochs}  &\colhead{Bandwidth} 
&\colhead{Typical Integration}\\
& & & & \colhead{ToAs} & &\colhead{(MHz)} &\colhead{Time (minutes)}}
\startdata
Mark III-A\tablenotemark{$\dagger$}(1410~MHz)& M3A-L & Arecibo & 1992 Jun$-$1993 Jan&  9&  9&  40&  47\\
Mark III-B\tablenotemark{$\ddagger$}(1410~MHz)& M3B-L & Arecibo & 1993 Jan$-$1994 Jan&  46&  46&  40&  47\\
Mark IV (1410~MHz)& M4-L & Arecibo & 1998 Jul$-$2004 May&  81&  81&  10&  58\\
Mark IV (2380~MHz)& M4-S & Arecibo & 1999 Oct$-$2004 May&  44&  44&  10&  29\\
Mark IV-O\tablenotemark{$\star$} (1410~MHz)& M4O-L &Arecibo & 2004 Jun$-$2005 Mar&  22&  16&  10&  60\\
Mark IV-O\tablenotemark{$\star$} (2380~MHz)& M4O-S & Arecibo & 2004 Jun$-$2005 Jan&  8&  7&  10&  30\\
ABPP (1410~MHz) & ABPP-L& Arecibo & 1998 Feb$-$2004 May&  98&  89&  56&  60\\
ABPP (2380~MHz) & ABPP-S& Arecibo & 1999 Dec$-$2004 May&  46&  46&  112&  30\\
ASP (1410~MHz) & ASP-L& Arecibo & 2005 Jan$-$2012 Jan&  990&  48&  64&  20\\
ASP (2350~MHz) & ASP-S& Arecibo & 2005 Jan$-$2012 Mar&  668&  41&  64&  20\\
GASP (800~MHz) & GASP-8& GBT & 2006 Mar$-$2011 Jan&  997&  41&  64&  20\\
GASP (1410~MHz) & GASP-L& GBT & 2006 Mar$-$2010 Jun&  863&  42&  64&  20\\
GUPPI (800~MHz) & GUPPI-8& GBT & 2010 Mar$-$2013 Oct&  3533&  49&  800&  20\\
GUPPI (1410~MHz)& GUPPI-L& GBT & 2010 Mar$-$2013 Nov&  4381&  64&  800&  20\\
PUPPI (1410~MHz)& PUPPI-L& Arecibo & 2012 Mar$-$2013 Nov&  1972&  26&  800&  20\\
PUPPI (2300~MHz)& PUPPI-S& Arecibo & 2012 Mar$-$2013 Nov&  992&  24&  800&  20
\enddata

\tablenotetext{*}{These short names are used in Figure \ref{fig:res} and \ref{fig:detres} and
Table \ref{tab:wrms}.}
\tablenotetext{$\dagger$}{Filter bank used a 78~$\mu$s time constant.}
\tablenotetext{$\ddagger$}{Filter bank used a 20~$\mu$s time constant.}
\tablenotetext{$\star$}{Here Mark IV-O stands for the recently processed Mark IV
data that partially overlap with ASP data.}

\end{deluxetable*}


The earliest observations (1992--1994) used the Princeton Mark~III
\citep{skn+92}, which collected dual-polarization data with a filter bank of 32
spectral channels each 1.25~MHz wide. 
Observations between 1998 and 2004 used the Princeton Mark~IV
\citep{sst+00} instrument and the ABPP (\citealt{bdz+97}) system  in parallel. 
The Mark~IV system collected 10-MHz passband data using 2-bit sampling. The
data were coherently dedispersed and folded at the pulse period offline.
The ABPP system sampled voltages with 2-bit resolution and filtered the passband 
into 32 spectral channels (1.75~MHz
per channel and 56~MHz in total for 1410-MHz band; 3.5~MHz per channel and 112~MHz in total for 2380-MHz band), and applied coherent dedispersion to each
channel using 3-bit coefficients.


From 2004 to 2011/12, pulsar data were collected with the Astronomical Signal
Processor (ASP; \citealt{dem07}) and its Green Bank counterpart GASP \citep{dem07}.
The (G)ASP systems recorded 8-bit sampled $\sim$64~MHz bandwidth data and
applyed real-time
coherent dedispersion and pulse period folding. The resulting data contain
2048-bin full-Stokes pulse profiles integrated over 1-3 minutes. 
When observing with ASP we used 16 channels each 4-MHz wide between 1440 and
1360~MHz, and 16 channels between 2318 and 2382~MHz. 
When observing with GASP we used 12 channels between 1386 and 1434~MHz, and 16
channels between 822 and 886~MHz.
The J1713+0747 ASP/GASP data were also reported in \citet{dfg+13}.

We started using 
the Green Bank Ultimate Pulsar Processing Instrument (GUPPI; \citealt{GUPPI}) for GBT 
observations in 2010 and its clone the Puerto-Rican Ultimate Pulsar Processing Instrument
(PUPPI) for Arecibo observations in 2012. 
GUPPI and PUPPI use 8-bit sampling in real-time coherent dedispersion and
pulse period folding mode and produce 2048-bin full-Stokes
pulse profiles integrated over 10 second intervals.
When observing with the 800-MHz receiver at the GBT, we use GUPPI to collect data from 62 spectral 
channels each 3.125~MHz wide, covering 724--918~MHz in frequency. With the L-band receiver, GUPPI
uses 58 spectral channels each 12.5~MHz wide, covering the 1150--1880~MHz band. 
When observing with the L-band receiver at Arecibo, we use PUPPI to collect data
from 1150--1765~MHz using 50 spectral channels each 12.5~MHz wide. When observing with the S band,
PUPPI takes data from 1770--1880 MHz and 2050--2405 MHz using 38 spectral
channels each 12.5~MHz wide.
The spectral bandwidth and resolution provided by GUPPI and PUPPI are crucial for resolving the pulse profile evolution in frequency described in Section \ref{sec:FD}.

\subsection{Arrival time calculations}

We combine the pulse times of arrival (TOAs) used in \citealt{sns+05} and those
of the later observations, for a data span of 21 years, with a
noticeable gap between 1994 and 1998, during the Arecibo upgrade.
We used daily averaged TOAs from \citealt{sns+05}, because the original 190~s integration TOAs are not accessible.
Data timestamps are derived from observatory masers and retrocorrected
to Universal Coordinated Time (UTC) via GPS and then further
corrected to the TT(BIPM) timescale using the 2012 version BIPM clock corrections with extrapolations to 2013.
The TOAs are measured from the observational data through a series of
steps. First the data are folded, as they were being taken, into pulse
profiles using an ephemeris known to be good enough for predicting the
pulse period for the duration of the observation. The folded
profiles from different frequency channels and sub-integrations are
often summed together to improve the signal-to-noise ratio of the profiles.  Orthogonal
polarizations are summed to produce a total-intensity profile.
The summed profiles are then compared with a well-measured standard
pulse profile from the appropriate frequency band. We employ
frequency-domain cross-correlation techniques \citep{tay92} to determine the phase of the pulse peak relative to the midpoint of the observation. The final TOA of a summed profile is then calculated by adding the mid-observation time and the product of pulse period and the measured peak phase.
The flux density of the pulsar in these observations can also be
measured by comparing the signal strength in the data with that of a
calibration observation taken right before or after the pulsar
observation in which a signal with known strength was injected. For
the post-upgrade Arecibo data and all the GBT data, the
flux density of the calibration signal is calibrated every month by
comparing it with an astronomical object of known and constant flux
density, in this case, the AGNs J1413+1509 and B1442+09.

\subsection{Instrumental offsets}
The telescopes and data acquisition systems introduce varying degrees of
computational and electronic delays into the measured TOAs.   Further, in many
cases, different standard pulse profiles were used to create TOAs from
different data acquisition systems.  As a result, the TOAs from different
systems differ by small time offsets.  When a transition is made from one
system to another at a telescope, typically data are collected in parallel
during a period of around a year, and those data are used to measure the
offset between data taken with the instruments.  Here we describe the measured
offsets in more detail.

There was no overlap between data taken with Mark III (through 1994) and
data taken with Mark IV and ABPP (beginning in 1998).  The offset between
these two systems was treated as a free parameter when fitting timing
solutions to the full data set.

Mark~IV and ABPP were used in parallel for J1713+0747 observations between
1998 and 2004. They collected data with different bandwidths (Table 1) and
were computed using different profile templates. To align the ABPP ToAs
with Mark IV, we fitted a phase offset between the 1410-MHz TOAs of the
two instruments, and found that ABPP ToAs trail those of Mark IV
by $0.46791\pm0.00009$ in pulse phase. In the timing modeling, we fix
the phase offsets between Mark~IV and ABPP to this value in
both 1410~MHz and 2300~MHz, and fit an extra time offset for 2300~MHz ABPP 
TOAs to account for any extra template misalignment.

Mark IV/ABPP and ASP were used in parallel for several epochs.  We fit
across the overlap data to determine the offset between the 1410 MHz
TOAs of these data sets to measure an offset of $2.33\pm0.10$~$\mu$s
(Mark IV trailing ASP). The offset between the 2300~MHz TOAs of Mark~IV 
and ASP is treated as a free parameter in the full timing solution.

For the offsets between ASP and PUPPI at Arecibo, and GASP and GUPPI
at the GBT, analysis of simultaneous observations of many pulsars,
including both pulsar signals and radiometer noise, were used to
measure very precise offsets between the instruments at each
observatory; details will be given in a forthcoming paper \citep{abb+15b}.  
Further, the same standard pulse profiles were used in
any given band.  Thus these data sets form a continuous 9-year
collection of TOAs with no arbitrary offsets.

Finally, the offset between the Arecibo 1410~MHz TOAs and GBT Bank 1410~MHz TOAs 
 was treated as a free parameter when fitting timing solutions to the full data set.  
The offset between the 800~MHz and 1410~MHz GBT TOAs and the offset between
the 1410~MHz and the 2300~MHz Arecibo TOAs are also fitted as free
parameters.  

The date span, number of observation epochs, and specifications of the
systems are listed in Table \ref{tab:obs}.

\section{Timing model}
\label{sec:model}
We employed the pulsar timing packages {\sc TEMPO}
\footnote{\url{http://tempo.sourceforge.net}} and {\sc TEMPO2} \citep{hem06} to model the TOAs. 

The rotation of the pulsar was modeled with a low-order polynomials expansion of spin frequency 
$\nu$ and $\dot{\nu}$ (Tables \ref{tab:par1} and \ref{tab:par2}) in order to account
for the pulsar's spin and spin down.

The pulsar's position ($\alpha$, $\delta$) and proper motion ($\mu_\alpha$, $
\mu_\delta$) on the sky and its parallax $\varpi$ were also measured through timing modeling. 
The distance inferred from our parallax
measurement is $D_{\rm PSR}=1.15\pm0.03$~kpc. This distance is consistent with 
the VLBA parallax distance of $1.05\pm0.06$~kpc \citep{cbv+09}.

We employed the \citet{dd86} (DD) model of binary motion to fit for the binary parameters, 
including the mass of the white dwarf ($M_{\rm c}$) measured
through Shapiro delay (Section \ref{sec:mass}),
the orbital period $P_{\rm b}$, angle of periastron $\omega$, time of
periastron passage $T_0$, projected semi-major axis $x$ and its change rate
$\dot{x}$ due to proper motion of the orbit. 
We observed an apparent $\dot{x}$ as the projection angle of the orbit
changed over time due to the perpendicular part of the binary's motion 
to our line of sight. This 
allowed us to determine the orientation of the orbit in the sky when combined
with the system's proper motion.
The orientation of the orbit in the sky is modeled by the
parameter $\Omega$, the position angle of the ascending node.
In {\sc TEMPO}, we grid search for the best $\Omega$ and then hold it fixed when fitting other
parameters (Table \ref{tab:par1}).
In the T2 model of {\sc TEMPO2} \citep{ehm06}, $\Omega$ is explicitly modeled and
fitted, while the changing of $x$, including the $\dot{x}$ caused by proper motion and the 
periodic changes due to orbital parallaxes of the Earth and the pulsar \citep{kop96}, are implicitly modeled and not fitted as a parameter
(Table \ref{tab:par2}).  

In our model, we fixed the orbit's periastron advance rate
$\dot{\omega}$ to the value inferred from GR and the best-fit
binary parameters (Table \ref{tab:par1}, \ref{tab:par2}). This was done by iteratively updating the values of $\dot{\omega}$ and
refitting for new binary parameters many times until the results converged.
Fitting for $\dot{\omega}$ as a free parameter
resulted in a best-fit value $\gtrsim 3\sigma$ away from the GR prediction.
This is likely because, in J1713+0747's nearly circular orbit, $\dot{\omega}$ is strongly 
covariant with the orbital period $P_{\rm d}$.
%

\begin{deluxetable*}{lccc}

\tabletypesize{\scriptsize}
\tablewidth{0pt}
\tablecaption{\label{tab:par1} Timing model parameters\tablenotemark{a} from {\it TEMPO}. }
\tablehead{ \colhead{Parameter}  &\colhead{EFAC and EQUAD}  &\colhead{With
Jitter Model}  &\colhead{Jitter and Red Noise Model}   }
\startdata
\textit{Measured Parameters}&  &  &  \\
R.A., $\alpha$ (J2000)&  17:13:49.5320251(5)&  17:13:49.5320248(7)&  17:13:49.5320252(8)\\
Decl., $\delta$ (J2000)&  7:47:37.506131(12)&  7:47:37.506155(19)&  7:47:37.50614(2)\\
Spin frequecy $\nu$~(s$^{-1}$)&  218.81184385472585(6)&  218.81184385472594(10)&  218.8118438547251(9)\\
Spin down rate $\dot{\nu}$ (s$^{-2}$)&  $-4.083889(4)\times10^{-16}$&  $-4.083894(7)\times10^{-16}$&  $-4.08382(5)\times10^{-16}$\\
Proper motion in $\alpha$, $\mu_{\alpha}=\dot{\alpha}\cos \delta$ (mas~yr$^{-1}$)&  4.9177(11)&  4.9179(18)&  4.917(2)\\
Proper motion in $\delta$, $\mu_{\delta}=\dot{\delta}$ (mas~yr$^{-1}$)& $-$3.917(2)&  $-$3.915(3)&  $-$3.913(4)\\
Parallax, $\varpi$ (mas)&  0.858(15)&  0.84(3)&  0.85(3)\\
Dispersion measure\tablenotemark{b} (pc~cm$^{-3}$)&  15.9700&  15.9700& 15.9700\\
Orbital period, $P_{\rm b}$ (day)&  67.82513682426(16)&  67.82513826935(19)&  67.82513826930(19)\\
Change rate of $P_{\rm b}$, $\dot{P}_{\rm b}$ ($10^{-12}$s~s$^{-1}$)&  0.23(12)&  0.41(16)&  0.44(17)\\
Eccentricity, $e$&  0.0000749394(3)&  0.0000749399(6)&  0.0000749402(6)\\
Time of periastron passage, $T_0$ (MJD)&  53761.03227(11)&  53761.0328(3)&  53761.0327(3)\\
Angle of periastron\tablenotemark{c}, $\omega$ (deg)&  176.1941(6)&  176.1967(15)&  176.1963(16)\\
Projected semi-major axis, $x$ (lt-s)&  32.34242243(5)&  32.34242188(14)&  32.34242188(14)\\
$\sin i$, where $i$ is the orbital inclination angle&  0.9672(11)&  0.951(4)&  0.951(4)\\
Companion mass, $M_c$ ($M_{\odot}$)&  0.233(4)&  0.287(13)&  0.286(13)\\
Apparent change rate of $x$, $\dot{x}$ (lt-s~s$^{-1}$)&  0.00637(7)&  0.00640(10)&  0.00645(11)\\
Profile frequency dependency parameter, FD1 &  $-$0.00016317(19)& $-$0.0001623(2)&  $-$0.00016(3)\\
Profile frequency dependency parameter, FD2 &  0.0001357(3)&  0.0001350(3)&  0.00014(3)\\
Profile frequency dependency parameter, FD3 &  $-$0.0000664(6)& $-$0.0000668(6)&  $-$0.000067(17)\\
Profile frequency dependency parameter, FD4 &  0.0000147(4)&  0.0000153(4)& 0.000015(5)\\[4pt]
\textit{Fixed Parameters}&  &  &  \\
Solar system ephemeris&  DE421&  DE421&  DE421\\
Reference epoch for $\alpha$, $\delta$, and $\nu$ (MJD)&  53729&  53729&  53729\\
Solar wind electron density $n_{\rm 0}$ (cm~$^{-3}$) & 0 & 0 & 0 \\
Rate of periastron advance, $\dot{\omega}$ (deg yr$^{-1}$)\tablenotemark{d}&  0.00020&  0.00024&  0.00024\\
Position angle of ascending node, $\Omega$ (deg)\tablenotemark{e}&  88.43&  88.43&  88.43\\
Red noise amplitude ($\mu$s~${\rm yr}^{1/2}$)&  --&  --&  0.025 \tablenotemark{f}\\
Red noise spectral index, $\gamma_{\rm red} $&  --&  --&  $-$2.92\\[4pt]
\textit{Derived Parameters}&  &  &  \\
Intrinsic period derivative, $\dot{P}_{\rm Int}$(s~s$^{-1}$)\tablenotemark{*}&  $8.966(12)\times10^{-21}$&  $8.98(2)\times10^{-21}$&  $8.97(2)\times10^{-21}$\\
Dipole magnetic field, $B$ (G)\tablenotemark{*}&  $2.0485(14)\times10^{8}$&  $2.050(3)\times10^{8}$&  $2.049(3)\times10^{8}$\\
Characteristic age, $\tau_c$ (year)\tablenotemark{*}& $8.076(11)\times10^{9}$& $8.07(2)\times10^{9}$&  $8.07(2)\times10^{9}$\\
Pulsar mass, $M_{\rm PSR}$ ($M_{\odot}$)&  0.97(3)&  1.32(11)&  1.31(11)
\enddata
\tablenotetext{a}{We used a modified {\it DD} binary model \citep{dd86} that
allows us to assume a position angle of ascending node ($\Omega$) and fit for
the apparent change rate of the projected semi-major axis ($\dot{x}$) due to
proper motion. Numbers in parentheses indicate the 1 $\sigma$ uncertainties on the last
digit(s). Uncertainties on parameters are estimated by the {\it TEMPO} program
using information in the covariance matrix.}
\tablenotetext{b}{The averaged DM value; see Section 3.2 and Figure 2 for more discussion.}
\tablenotetext{c}{See Figure 2 of \citealt{sns+05} for definition.}
\tablenotetext{d}{The rate of periastron advance was not fitted but fixed to the GR value
because it is highly co-variant with the orbital period. }
\tablenotetext{e}{We optimized $\Omega$ using a grid search and held it fix to the value that
minimized $\chi^2$.}
\tablenotetext{f}{The value corresponds to $8.7\times10^{-15}$ in the dimensionless strain amplitude unit.}
\tablenotetext{*}{These parameters are corrected for Shklovskii effect and
Galactic differential accelerations.}

\end{deluxetable*}


\begin{deluxetable*}{lccc}

\tabletypesize{\scriptsize}
\tablewidth{0pt}
\tablecaption{\label{tab:par2} Timing model parameters\tablenotemark{a} from
{\sc TEMPO2}. }
\tablehead{ \colhead{Parameter}  &\colhead{EFAC \& EQUAD}  &\colhead{With Jitter Model}  &\colhead{Jitter \& Red Noise Model}   }
\startdata
\textit{Measured Parameters}&  &  &  \\
R. A., $\alpha$ (J2000)&  17:13:49.5320254(5)&  17:13:49.5320247(7)&  17:13:49.5320261(10)\\
Decl., $\delta$ (J2000)&  7:47:37.506130(13)&  7:47:37.506130(19)&  7:47:37.50615(3)\\
Spin frequecy $\nu$~(s$^{-1}$)&  218.81184385472573(6)&  218.81184385472589(10)&  218.8118438547255(5)\\
Spin down rate $\dot{\nu}$ (s$^{-2}$)&  $-4.083883(5)\times10^{-16}$&  $-4.083892(7)\times10^{-16}$&  $-4.08386(5)\times10^{-16}$\\
Proper motion in $\alpha$, $\mu_{\alpha}=\dot{\alpha}\cos \delta$ (mas~yr$^{-1}$)&  4.9161(12)&  4.9181(18)&  4.915(3)\\
Proper motion in $\delta$, $\mu_{\delta}=\dot{\delta}$ (mas~yr$^{-1}$)& $-$3.915(2)&  $-$3.910(3)&  $-$3.914(5)\\
Parallax, $\varpi$ (mas)&  0.872(16)&  0.83(3)&  0.87(3)\\
Dispersion measure\tablenotemark{b} (pc~cm$^{-3}$)&  15.9700&  15.9700&  15.9700\\
Orbital period, $P_{\rm b}$ (day)\tablenotemark{c}&  67.8251365449(12)&  67.8251383194(16)&  67.8251383185(17)\\
Change rate of $P_{\rm b}$, $\dot{P}_{\rm b}$ ($10^{-12}$ s~s$^{-1}$)&  0.19(13)&  0.39(16)&  0.36(17)\\
Eccentricity, $e$&  0.0000749395(3)&  0.0000749400(6)&  0.0000749402(6)\\
Time of periastron passage, $T_0$ (MJD)&  53761.03208(11)&  53761.0327(3)&  53761.0328(3)\\
Angle of periastron, $\omega$ (deg)&  176.1930(6)&  176.1963(15)&  176.1966(14)\\
Projected semi-major axis, $x$ (lt-s)&  32.34242258(5)&  32.34242189(13)&  32.34242187(13)\\
Orbital inclination, $i$ (deg)&  76.1(3)&  71.9(7)&  71.9(7)\\
Companion mass, $M_c$ ($M_{\odot}$)&  0.222(4)&  0.286(13)&  0.286(12)\\
Position angle of ascending node, $\Omega$ (deg)&  74.3(14)&  89.6(20)&  88(2)\\
Profile frequency dependency parameter, FD1 &  $-$0.0001634(2)& $-$0.0001628(2)&  $-$0.0001628(2)\\
Profile frequency dependency parameter, FD2 &  0.0001358(3)&  0.0001355(3)&  0.0001355(3)\\
Profile frequency dependency parameter, FD3 &  $-$0.0000658(6)& $-$0.0000671(6)&  $-$0.0000672(6)\\
Profile frequency dependency parameter, FD4 &  0.0000141(4)&  0.0000154(4)& 0.0000155(4)\\[4pt]
\textit{Fixed Parameters}&  &  &  \\
Solar system ephemeris&  DE421&  DE421&  DE421\\
Reference epoch for $\alpha$, $\delta$, and $\nu$ (MJD)&  53729&  53729&  53729\\
Solar wind electron density $n_{\rm 0}$ (cm~$^{-3}$)& 0 & 0 & 0 \\
Rate of periastron advance, $\dot{\omega}$ (deg yr$^{-1}$)\tablenotemark{d}&  0.00019&  0.00024&  0.00024\\
Red noise amplitude ($\mu$s~${\rm yr}^{1/2}$)&  --&  --&  0.025 \tablenotemark{e}\\
Red noise spectral index, $\gamma_{\rm red}$&  --&  --& $-$2.92\\[4pt]
\textit{Derived Parameters}&  &  &  \\
Intrinsic period derivative, $\dot{P}_{\rm Int}$(s~s$^{-1}$)\tablenotemark{*}&  $8.957(13)\times10^{-21}$&  $8.98(2)\times10^{-21}$&  $8.96(2)\times10^{-21}$\\
Dipole magnetic field, $B$ (G)\tablenotemark{*}&  $2.0473(15)\times10^{8}$&  $2.050(3)\times10^{8}$&  $2.048(3)\times10^{8}$\\
Characteristic age, $\tau_c$ (year)\tablenotemark{*}&  $8.085(12)\times10^{9}$& $8.06(2)\times10^{9}$&  $8.08(2)\times10^{9}$\\
Pulsar mass, $M_{\rm PSR}$ ($M_{\odot}$)&  0.90(3)&  1.31(11)&  1.31(11)
\enddata
\tablenotetext{a}{We used {\sc TEMPO2}'s {\it T2} binary model, which
implicitly account for the changes of the projected semi-major axis, including
$\dot{x}$ due to proper motion of the binary (this allows us to fit for the
position angle of ascending node, $\Omega$) and the changes due to the orbital
parallaxes of the earth and the pulsar \citep{kop96, ehm06}. 
Numbers in parentheses indicate the 1 $\sigma$ uncertainties on the last digit(s).  
Uncertainties on parameters are estimated by the {\sc TEMPO2} program using information in the covariance matrix.
These uncertainties are consistent with the MCMC results using the full non-linear timing model (Section
\ref{sec:noise}), see Figure \ref{fig:masses} for examples.
}
\tablenotetext{b}{The averaged DM value; see Section 3.2 and Figure 2 for more discussion.}
\tablenotetext{c}{{\sc TEMPO2}'s {\it T2} model reports the orbital period
after correcting for the changes in periastron angle $\omega$ due to
proper motion and orbital parallaxes. {\sc TEMPO} {\it DD} model does not account for this and
therefore reports a slightly different value.}
\tablenotetext{d}{The rate of periastron advance was not fitted but fixed to the GR value
because it is highly co-variant with the orbital period.}
\tablenotetext{e}{This value corresponds to $8.7\times10^{-15}$ in
the dimensionless strain amplitude unit.}
\tablenotetext{*}{These parameters are corrected for Shklovskii effect and
Galactic differential accelerations.}

\end{deluxetable*}


Compared with previous timing efforts, we detected, for the first time, an
apparent change in the binary period $\dot{P}_{\rm b}$, which we attribute to
the motion of the binary system relative to the Sun. This is described in Section \ref{sec:obdecay}.    

The DMX model was used to fit dispersion measure (DM) variations caused by
changes in the ISM along the line of sight (see Section \ref{sec:dmx} for details). The FD model was used to model profile
evolution in frequency (see Section \ref{sec:FD} for details). 

In order to account for unknown systematics in TOAs from different
instruments, and observation-correlated noise such as pulse jitter noise from pulsar
emission process, we employed a general noise model that parameterizes both uncorrelated and
correlated noise. The noise modeling is discussed in Section \ref{sec:noise} and the
noise model parameters are listed in Table \ref{tab:wrms}. 

We used the JPL DE421 solar system ephemeris \citep{fwb09} to remove pulse
time-of-flight variation within the solar system. This ephemeris is oriented
to the International Celestial Reference Frame (ICRF) and thus our astrometric
results are also given in the ICRF frame. We note parenthetically that a
previous generation ephemeris, DE405, gave nearly identical but marginally
better timing fits ($\Delta\chi^2\sim6$ for 14528 dof in both
{\sc TEMPO} and {\sc TEMPO2}). 

The timing parameters and uncertainties are calculated using a generalized least
square (GLS) approach available in the {\sc TEMPO} and {\sc TEMPO2} software packages. 
Furthermore we also run a Markov Chain Monte Carlo (MCMC) \citep[similar to][]{lah+14, vv14} 
analysis that simultaneously 
includes the noise parameters (See Section \ref{sec:noise}) and the nonlinear timing model. 
As shown in Table \ref{tab:par2} and Figure \ref{fig:masses}, the results from this analysis
are very consistent with the GLS approach, indicating that the assumption of linearity holds
over the full timing parameter range.

Using a new noise modeling technique, we detected 
a significant red noise signal that could be the same ``timing noise''
described in \citet{sns+05}.
We also detected for the first time a change in the observed orbit period.
The new timing model parameters (Table \ref{tab:par1} and \ref{tab:par2}) changed slightly 
from those in \citet{sns+05} but are consistent within their reported uncertainties.

\subsection{Noise model}
\label{sec:noise}

\begin{figure*}[!ht]
\centering
\includegraphics[scale=1.0]{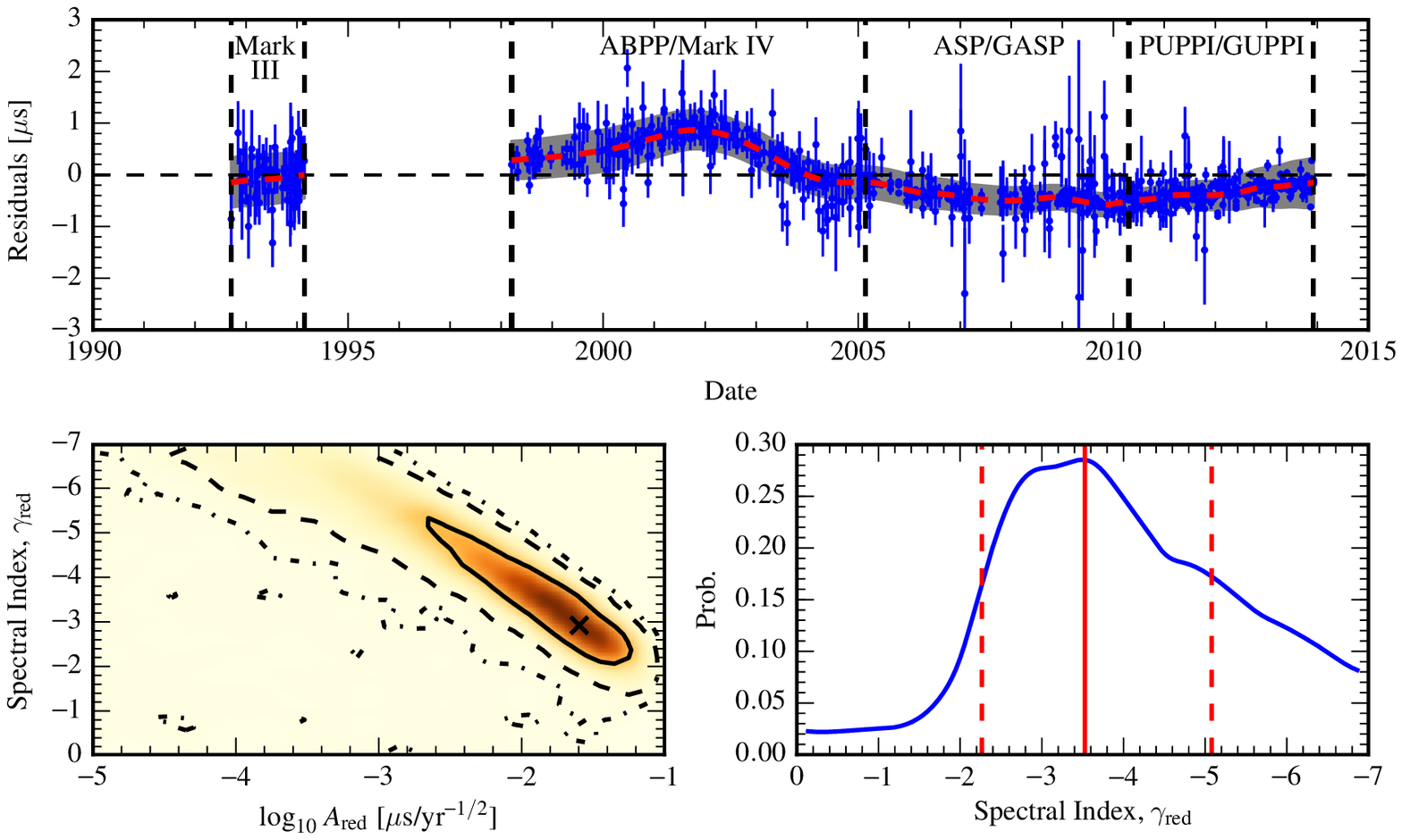} \\ 
\caption{\label{fig:res} Top panel: daily-averaged residuals of J1713+0747 produced from a GLS that includes a 
full noise model. The red dashed line and the gray shaded area show the maximum likelihood red noise
realization and one-sigma uncertainty. The thick dashed vertical lines separate out various generations
of backends used at both AO and GBT.
Bottom-left panel: two-dimensional marginalized posterior probability plot of red
noise spectral index vs. logarithm of the red noise amplitude where the solid, dashed and
dashed-dotted lines represent the 50, 90, and 95 percent credible regions. The
``$x$'' denotes
the maximum likelihood value of the spectral index and amplitude.
Bottom-right panel: one-dimensional marginalized posterior probability for the
red noise spectral index where the solid line denotes the maximum marginalized a-posteriori
value and the dashed lines denote the 68\% credible interval. Note that the
1D maximum marginalized posterior spectral index usually differs slightly from the
global maximum likelihood value ($-2.92$) due to correlations with other
parameters.}
\end{figure*} 
The noise model used in this analysis is a parameterized model that is a function of several unknown quantities describing both correlated and uncorrelated noise sources. Uncorrelated noise is 
independent from one TOA to another, while the correlated noise is not. 
For instance, the template matching error mostly due to radiometer noise are
uncorrelated in time, but the pulse jitter noise \citep{cs10}, which affects
the multi-frequency TOAs measured simultaneously across the band, is correlated in time.
There is also time-correlated noise, such as red timing noise that is
correlated from epoch to epoch. Among the various types of noise only the
template matching error $\sigma$ can be estimated when we compute the TOAs.
Other sources of noise must be modeled separately. The solution to this
problem has been discussed extensively in \citet{vl13, ell13, vhv14a, vhv14,
abb+14} and \citet{ell14}. In this section, we summarize the noise model used in this analysis \citep[see e.g.][for more details]{abb+15b}.

\begin{figure*}[!ht]
\centering
\includegraphics[scale=1.0]{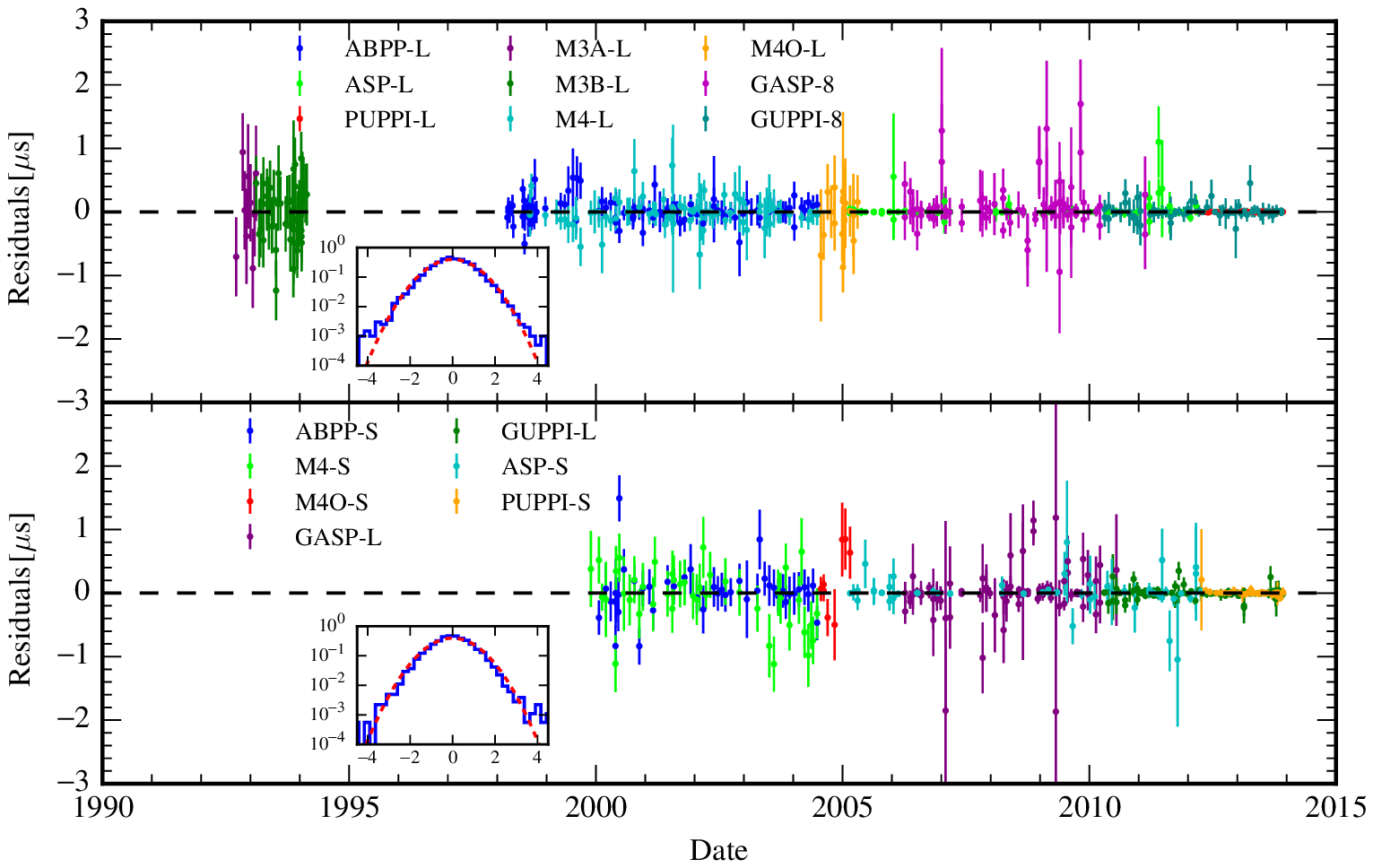} \\ 
\caption{\label{fig:detres} Top panel: low frequency daily averaged residuals of J1713+0747 with maximum likelihood
jitter and red noise realization subtracted out. Bottom panel: high frequency
daily averaged residuals of J1713+0747 with maximum likelihood jitter and red
noise realization subtracted out. The insets in both top and bottom panels
show the histogram of the weighted residuals (weighted by EFAC and EQUAD
corrected TOA uncertainties) in logarithmic scale in blue curves, along with a
zero-mean unit-variance Gaussian distribution marked as red dashed curves.}
\end{figure*} 

To model uncorrelated noise, we use the standard EFAC and EQUAD parameters for each
backend/receiver system (e.g. PUPPI backend with L-band receiver). These parameters
simply re-scale the original TOA uncertainties
\be
\sigma_{i,k} \rightarrow  E_{k} \left( \sigma_{i,k}^{2} + Q_{k}^{2}  \right)^{1/2},
\ee
where $E_{k}$ and $Q_{k}$ denote the EFAC and EQUAD parameters, respectively, and the subscript
$i$ is the TOA number and the subscript $k$ denotes the backend/receiver system.

To model correlated noise we use the new ECORR parameter and a power-law red noise spectrum.
The ECORR term describes short timescale noise that is completely correlated
for all TOAs in a given observation
but completely \emph{uncorrelated} between observations. This term could be described as pulse phase jitter \citep{cs10}
but could also have other components. The ECORR term manifests itself as a block diagonal term in the noise covariance
matrix where the size of the blocks is equal to the number of TOAs in a given
observation. The exact details of the implementation are described in
\cite{abb+15b}; however, the term essentially acts as a
observation-to-observation variance. 
Lastly, we model the red noise as a stationary gaussian process that is parameterized by a power spectrum of the form
\be
P(f) = A_{\rm red}^2\left( \frac{f}{f_{\rm yr}} \right)^{\gamma_{\rm red}},
\ee
where $A_{\rm red}$ is the amplitude of the red noise in $\mu$s~${\rm yr}^{1/2}$, $\gamma_{\rm red}$ is the spectral index.

These noise parameters are included in a joint likelihood that contains all timing model parameters. For the purposes of noise modeling, we analytically marginalize over the linear timing model parameters and explore the space of noise parameters via MCMC. We then use the MCMC results to determine the maximum likelihood noise parameters which are subsequently used as inputs to \textsc{TEMPO}/\textsc{TEMPO2} GLS fitting routines. 
In our noise model we include EFAC, EQUAD, and ECORR parameters
for data collected by different backend and receiver systems. However, we do not model 
EFAC and ECORR of the Mark~III, Mark~IV, and ABPP data because there were not
enough TOAs in the legacy data set to constrain both EFAC and EQUAD. Furthermore,
there was only one TOA per observation so we cannot constrain the observation-correlated
noise modeled by ECORR. Instead, we set EFAC values to 1
and ECORR to 0 for these data sets, and use only EQUAD to model the white
noise in these observations. 
The maximum likelihood values of the white noise parameters are presented in Table \ref{tab:wrms}.

\begin{deluxetable}{lcccccc}

\tabletypesize{\footnotesize}
\tablewidth{0pt}
\tablecaption{\label{tab:wrms} Noise parameters\tablenotemark{a} and residual rms in $\mu$s. }
\tablehead{\colhead{Backends}  &\colhead{$\bar{\sigma}$\tablenotemark{b}}  &\colhead{EFAC}  &\colhead{EQUAD}  &\colhead{ECORR}  &\colhead{WRMS\tablenotemark{c}}  &\colhead{AWRMS\tablenotemark{d}}   }

\startdata
All&  0.927&  ...&  ...&  ...&  0.246&  0.092\\
M3A-L&  0.267&  ...&  0.599&  ...&  0.589&  0.588\\
M3B-L&  0.167&  ...&  0.412&  ...&  0.434&  0.432\\
M4-L&  0.172&  ...&  0.153&  ...&  0.365&  0.146\\
M4-S&  0.183&  ...&  0.357&  ...&  0.668&  0.416\\
M4O-L&  0.416&  ...&  0.315&  ...&  0.324&  0.112\\
M4O-S&  0.355&  ...&  0.008&  ...&  0.277&  0.141\\
ABPP-L&  0.106&  ...&  0.154&  ...&  0.288&  0.067\\
ABPP-S&  0.134&  ...&  0.260&  ...&  0.464&  0.303\\
ASP-L&  0.512&  0.979&  0.035&  0.105&  0.222&  0.073\\
ASP-S&  0.631&  1.149&  0.004&  0.127&  0.257&  0.115\\
GASP-8&  1.391&  1.178&  0.000&  0.023&  0.589&  0.098\\
GASP-L&  0.966&  1.128&  0.040&  0.037&  0.288&  0.080\\
GUPPI-8&  1.550&  1.052&  0.086&  0.204&  0.657&  0.156\\
GUPPI-L&  0.855&  1.204&  0.025&  0.054&  0.266&  0.046\\
PUPPI-L&  0.303&  1.160&  0.001&  0.094&  0.145&  0.075\\
PUPPI-S&  0.653&  1.050&  0.058&  0.114&  0.189&  0.080
\enddata

\tablenotetext{a}{The unmodeled EFAC and ECORR values default to 1 and 0,
respectively.}
\tablenotetext{b}{The averaged TOA template matching errors
$\left<\sigma\right>$.}
\tablenotetext{c}{Here WRMS is defined as $[(\sum w_i\msR_i^2/\sum w_i) -
\bar{\msR}^2]^{1/2}$, where $\msR_i$ is the timing residual of TOA $i$,
$w_i=1/\sigma_i^2$ is the weight determined by the TOA errors including EFAC
and EQUAD, and $\bar{\msR} = (\sum w_i\msR_i)/\sum w_i$ is the weighted mean
of the residuals after removing a red-noise model (as in Figure
\ref{fig:res}).}
\tablenotetext{d}{AWRMS stands for the weighted RMS of epoch-averaged
residuals.}

\end{deluxetable}


\citet{sc12} studied the pulse arrival times from a single long exposure of
PSR~J1713+0747, and found that this pulsar's single pulses showed random jitter of
$\simeq26~\mu$s. 
A similar result of $\simeq27~\mu$s was found by \citet{dlc+14} from a more
recent study using a 24 hr continuous observation of PSR~J1713+0747 conducted with major telescopes around the globe.
Therefore, by averaging many pulses collected in the typical
$\sim20$~minutes NANOGrav observation, one expects $\sim27\mu{\rm s}/\sqrt{1200\nu}=51$~ns of jitter noise. 
Tables \ref{tab:par1} and \ref{tab:par2} show the best-fit timing parameters before and
after we applied our noise model to the data. It is clear that the jitter-like
observation-correlated noise affected the arrival time of the pulses, such that 
some timing parameters changed significantly after including the jitter model.
The optimal jitter parameters (ECORR, as shown in Table \ref{tab:wrms}) from
our noise modeling are mostly consistent with the prediction from
\citet{sc12}, with some of them being higher. This could be due to the
covariance between the jitter parameters and the EQUAD parameters.

In Figure \ref{fig:res} we show the red noise realization based on our best
noise model (Table \ref{tab:wrms}) and compare it to the post-fit residuals of
a {\sc TEMPO} GLS fit. The bottom
panels of Figure \ref{fig:res} show the one- and two-dimensional posterior probability plots
of the red noise. This noise model describes the data well as we can see in
Figure \ref{fig:detres} in which the maximum likelihood realizations of red
and jitter noise are subtracted out. We see from the figure that both the high
and low frequency residuals (with red and jitter noise realizations
subtracted) are white (described by our EFAC and EQUAD parameters) and the
weighted residuals follow a zero-mean unit-variance Gaussian distribution. We
do note that the normalized residuals do not seem to follow exactly the gaussian 
distribution outside of the 3-sigma range, this affects only $\lesssim0.3$\% of the TOAs 
and will not significantly affect the results presented here. 

Red noise was previously reported for PSR~J1713+0747 by \citet{sns+05}, who
modeled it using an eighth-order polynomial. 
The largest residual in the
present data set appears in the time period 1999 to 2005 (Figure
\ref{fig:res}). This is when dual-frequency observations begin, and it is
likely that the red noise model is absorbing unmodeled DM variations in
singe-frequency data collected before 1999, thus making physical
interpretations of red noise difficult.

We reanalyzed the \citealt{sns+05} data sets (Mark~III, Mark~IV, and ABPP data)
with the new red noise modeling technique, and show that the timing
results from the new method are mostly consistent with those from \citealt{sns+05},
except for $\alpha$, $\delta$, $\nu$ and $\dot{\nu}$, which probably changed due
to the new red noise model, and Keplerian parameters, which probably changed due 
to the fact that we used {\sc TEMPO2}'s T2 model instead of {\sc TEMPO}'s DD
model used by \citet{sns+05}. We fit the Keplerian and post-Keplerian 
parameters simultaneously using the T2 model, whereas, \citet{sns+05} fit only for the
Keplerian parameters while holding the post-Keplerian parameters to their
best-fit values using the DD model.
 Therefore, the uncertainties reported in
\citealt{sns+05} do not reflect the covariance between the two sets of
parameters but ours do.

The red noise signal found by our noise model applied to the \cite{sns+05} data set is consistent with that found 
in the 21 year data set, both in terms of the noise parameters (Table
\ref{tab:par1}, \ref{tab:par2}, and \ref{tab:sns05}), and in terms of the shapes of the red noise
realization (Figure \ref{fig:res}), while the red noise modeled by high-order frequency
polynomials varies significantly depending on the order of the polynomial 
or and the observation time span.

\begin{deluxetable*}{lcc}

\tabletypesize{\scriptsize}
\tablewidth{0pt}
\tablecaption{\label{tab:sns05} The Timing Results from 
\citealt{sns+05} and from a Re-analysis of the \cite{sns+05} Data set
Using New Red Noise Analysis Technique.}
\tablehead{ \colhead{Parameter}  &\colhead{\citealt{sns+05}}  &\colhead{Red Noise Model}\tablenotemark{a}  }
\startdata
R. A., $\alpha$ (J2000)&  17:13:49.5305335(6)&  17:13:49.5305321(6)\\
Decl., $\delta$ (J2000)&  7:47:37.52636(2)&  7:47:37.52626(2)\\
Spin frequecy $\nu$~(s$^{-1}$)&  218.8118439157321(3)& 218.811843915731(1)\\
Spin down rate $\dot{\nu}$ (s$^{-2}$)&  $-4.0835(2)\times10^{-16}$&  $-4.0836(1)\times10^{-16}$\\
Proper motion in $\alpha$, $\mu_{\alpha}=\dot{\alpha}\cos \delta$ (mas~yr$^{-1}$)&  4.917(4)&  4.917(4)\\
Proper motion in $\delta$, $\mu_{\delta}=\dot{\delta}$ (mas~yr$^{-1}$)& $-$3.93(1)&  $-$3.93(1)\\
Parallax, $\varpi$ (mas)&  0.89(8)&  0.84(4)\\
Dispersion measure (pc~cm$^{-3}$)&  15.9960&  15.9940\\
Orbital period, $P_{\rm b}$ (day)&  67.8251298718(5)\tablenotemark{b}  &  67.825129921(4)\\
Change rate of $P_{\rm b}$, $\dot{P}_{\rm b}$ ($10^{-12}$s~s$^{-1}$)& 0.0(6)&  $-$0.2(7)\\
Eccentricity, $e$&  0.000074940(1)\tablenotemark{b}  &  0.000074940(1)\\
Time of periastron passage, $T_0$ (MJD)&  51997.5784(2) \tablenotemark{b} & 51997.5790(6)\\
Angle of periastron, $\omega$ (deg)&  176.192(1)\tablenotemark{b} &  176.195(3)\\
Projected semi-major axis, $x$ (lt-s)&  32.34242099(2)\tablenotemark{b} &  32.3424218(3)\\
Cosine of inclination, $\cos i$&  0.31(3)&  0.32(2)\\
Companion mass, $M_{\rm c}$ ($M_{\odot}$)&  0.28(3)&  0.30(3)\\
Position angle of ascending node, $\Omega$ (deg)&  87(6)&  89(4)\\
Solar system ephemeris&  DE405&  DE405\\
Reference epoch for $\alpha$, $\delta$, and $\nu$ (MJD)&  52000&  52000\\
Pulsar mass, $M_{\rm PSR}$ ($M_{\odot}$)&  1.3(2)&  1.4(2)\\
Red noise amplitude ($\mu$s~${\rm yr}^{1/2}$)&  --&  0.004\\
Red noise spectral index&  --&  5.14\\
\enddata
\tablenotetext{a}{We used {\sc TEMPO2}'s {\it T2} binary model, which models
the Keplerian ($P_{\rm b}$, $x$, $e$, $T_0$, and $\omega$) and
post-Keplerian orbital elements ($\cos i$, $\Omega$, and $m_2$ ) simultaneously.}
\tablenotetext{b}{\citealt{sns+05} uses {\sc TEMPO}'s DD model and reports the uncertainties of the Keplerian
parameters with the post-Keplerian ones fixed to their bestfit values. }
\end{deluxetable*}

\subsection{DM variation}
\label{sec:dmx}
The DM of a pulsar reflects the number of free electrons between
the pulsar and the telescopes and it varies because
our sight-line through the turbulent ISM and solar wind is changing as the
pulsar, the Sun, the Earth, and the ISM all move with respect to each other.
DM variation can affect the timing of high-precision pulsars significantly.

We fit simultaneously with other parameters the time-varying DM using the {\it
DMX} model in {\sc TEMPO}.
This model fits independent DM values for TOA groups taken within 14 day
intervals, except for the L-band-only Mark~III TOAs. We grouped the Mark~III
TOAs together as a single group, because their frequency resolution and timing
precision are not sufficient for measuring epoch-to-epoch DM changes.

We turned off the solar wind model for \textsc{TEMPO} and \textsc{TEMPO2} by
setting the solar wind electron density (at 1~AU from the Sun) parameter
$n_{\rm 0}$ to 0~cm$^{-3}$ (the default value is 10~cm$^{-3}$), and used the
DMX model to model all DM variations including contribution from solar wind.

\begin{figure}
\includegraphics[width=3.4in]{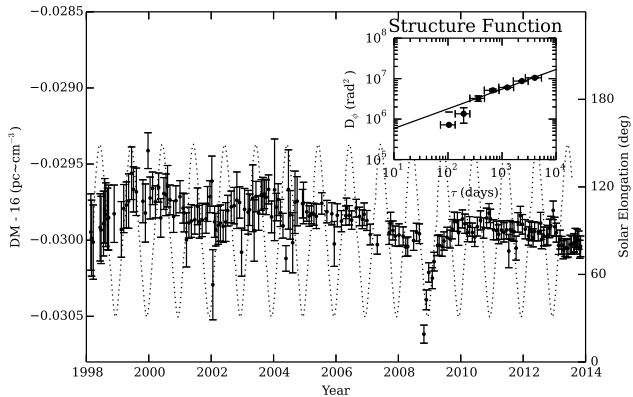} \\ 
\caption {\label{fig:dmx} The plot shows 16-years DM variation of PSR~J1713+0747. The dotted line shows the Solar
elongation of the pulsar. The subplot shows the structure
function (error bars) and its a power law fit (solid line). The best-fit power law index is
0.49(5), different from the value of $5/3$ expected from a ``pure'' Kolmogorov medium. } 
\end{figure}

Figure \ref{fig:dmx} shows the measured DM variation of PSR~J1713+0747.
The sudden dip and recovery of DM around 2008 (MJD 54800) is 
due to changes either in the ISM or in the solar wind. This DM dip is also
observed independently by the Parkes observatory \citep{kcs+13} and the
European Pulsar Timing Array (G. Desvignes et.~al.\ 2015, in preparation).

Spectrum analysis of the time variation of flux, pulse arrival phase, and DM have 
been employed to study the turbulent nature of the ISM \citep[e.g.][]{cpl86, rl90}.
It has been shown that DM variations of some pulsars are consistent with
those expected from an ISM characterized by a Kolmogorov turbulence spectrum
\citep{cwd+90, ric90, ktr94, yhc+07, kcs+13, fst14}. One can calculate the 
structure function of the varying DM: 
\begin{equation}
D_{\phi}(\tau)=\left(\frac{2\pi K}{f^2}\right)\langle [DM(t+\tau)-DM(t)]^2\rangle, 
\end{equation}
where $\tau$ 
is a given time delay, $K=4.148\times10^3$~MHz$^2$~pc$^{-1}$~cm$^3$~s, and $f$ is 
the observing frequency in MHz. We expect, under the simplest assumptions, 
this function to follow a Kolmogorov power law $D_{\phi}(\tau)=(\tau/\tau_0)^{\beta -2}$, 
where $\beta=11/3$ and $\tau_0$ is a characteristic time scale related to 
the inner scale of the turbulence. The pulsars with DM variations that fit this
theory generally have large DM variations on timescale of 
years. However, PSR J1713+0747 does not show significant long-term DM variation 
(Figure \ref{fig:dmx}). Conversely, it went through a steep drop and recovery 
around 2008. If such rapid DM changes are the result of variations in the ISM along
light of sight, such ISM variations do not fit the general characteristics of
a Kolmogorov medium. 


\subsection{Pulsar spin irregularity}
\label{sec:spin}

The term ``timing noise'' in pulsar timing generally refers to the non-white
noise left in the timing residuals.
An important contribution to timing noise is expected to come from the pulsar's spin
irregularity, i.e., its long-term deviation from a simple linear slow down. 
Spin irregularity is often significant in younger pulsars, and 
may be modeled with high-order frequency polynomials (such as $\ddot{\nu}$, where $\nu$ is the pulsar's spin frequency). 
Potential causes of irregular spin behavior include unresolved
micro-glitches, internal superfluid turbulence, magnetosphere variations, or external torques caused by matter surrounding the pulsar \citep{hlk10, ymh+13, ml14}.
These mechanisms could lead to accumulative random perturbations in the 
pulsar's pulse phase, spin rate, or spin-down rate. 
\citet{sc10} pointed out that one could model these types of timing noise using random walks.
Random walks in phase (RW$_0$) would grow over time ($T$)
proportionally to $T^{1/2}$, random walks in $\nu$ grow proportionally to
$T^{3/2}$, random walks in
$\dot{\nu}$ grow proportionally to $T^{5/2}$.
Such spin noise would likely have a steep power spectrum with more power in
the lower frequencies. This
is considered as one of the main sources of ``red'' noise in pulsar timing.

The timing noise of radio pulsars has been studied by
\citet{ch80,cd85,antt94,dmhd95, mtem97}, and later by \citet{hlk10} and
\citet{sc10} with large samples. 
\citet{mtem97} adopted a generalized Allen Variance (traditionally used in
measuring clock stability) to characterize the timing instability of pulsars:
\begin{equation}
\label{eq:sigmaz}
\sigma_z(\tau) = \frac{\tau^2}{2\sqrt{5}}\langle c^2 \rangle^{1/2},
\end{equation}
where $\langle c^2\rangle$ denotes the sum of squares of the cubic
terms fitted to segments of length $\tau$. 
\citet{hlk10} found a best-fit scaling model of $\sigma_z({\rm 10~yr})$ 
from a large sample of pulsars, including canonical pulsars (CPs) and MSPs:
\begin{equation}
\label{eq:hlk10}
\log_{10}[\sigma_z({\rm 10~yr})] =
-1.37\log_{10}[\nu^{0.29}|\dot{\nu}|^{0.55}]+0.52,
\end{equation} 
where $\nu$, $\dot{\nu}$ are the pulsar's spin and spin-down rate.
We find that \citet{hlk10}s scaling model ($\sigma^{\rm model}_{z, \rm
10~yr}\simeq1\times10^{-12}$) over-predicted $\sigma^{\rm measured}_{z, \rm
10~yr}=5\times10^{-16}$ for PSR J1713+0747 by more than three orders of magnitude. 

\citet{ch80} defined a different timing noise characteristic $\sigma^2_{\rm
TN,2}$ based on the root mean square of residuals $\sigma^2_{\msR,2}$ from a
timing fit that does not include any higher order spin parameters like
$\ddot{\nu}$. 
The timing noise term is related to $\sigma^2_{\msR,2}$:
\begin{equation}
\sigma^2_{\msR,2}(T) = \sigma^2_{\rm TN,2}(T) + \sigma^2_W, 
\end{equation}
where $\sigma^2_W$ is a time-independent term caused by white 
noise in the data.
In this definition, timing noise $\sigma^2_{\rm TN,2}(T)$ grows bigger over
time while white noise stays constant.  

\citet{sc10} studied the $\sigma^2_{\rm TN,2}$ from a large sample of CPs and
MSPs. They found a scaling model:
\begin{equation}
\label{eq:sc10}
\ln(\hat{\sigma}_{\rm TN,2}) = 1.6 - 1.4\ln(\nu) +
1.1\ln|\dot{\nu}_{-15}|+2\ln(T_{\rm yr}),
\end{equation}
where $\dot{\nu}_{-15}$ is $\dot{\nu}$ in units of $10^{-15}$s$^{-2}$, and $T_{\rm yr}$
is the observation time span in years.
This scaling model predicts that, for 21-year timing of PSR~J1713+0747, the
residual RMS without removing timing noise $\sigma^2_{\rm TN,2}$ would be
$\sim400$~ns. The measured RMS of the red noise residual 
$\sigma^2_{\msR,RN}=364$~ns, is consistent with the extrapolation
from \citet{sc10}.  
The best-fit scaling law also indicates that the residuals of the
sampled pulsars $\hat{\sigma}_{\rm TN,2}$ seem to grow linearly with $T_{\rm yr}^2$. 
If the timing noise of the sampled pulsars is due to the accumulation of 
spin noise, and the spin noise is caused by the same physical processes,
then this RMS growth rate would imply that the spin noise of pulsars has a
frequency power spectrum of power-law index $\gamma_{\rm red}\simeq -5$. This 
spectral index is consistent with the $\gamma_{\rm red}$
from our noise model. This can be seen in the bottom right plot of Figure \ref{fig:res} 
by noting that a spectral index of $-5$ is consistent with the posterior at the one-sigma
level. 

It is inconclusive whether or not the observed red noise can be interpreted as pulsar spin irregularity.
Other sources of noise also could have contributed significantly.
If we do assume that they are from spin irregularity, 
the estimated maximum likelihood red noise spectral index of $\sim-3$ 
favors that the pulsar spin irregularities come from
random walks in either spin phase or spin rate, although other explanations cannot be ruled out due to the
substantial uncertainty on the red noise spectral index (Figure \ref{fig:res}).

Finally, \citet{sc10} showed that the significance of timing noise coming from
gravitational wave (GW) background could be estimated as
$\sigma_{\rm GW,2} \approx$~1.3$A_0(T_{\rm yr})^{5/3}$~ns, where $A_0$ is the
characteristic strain at $f=1$~year$^{-1}$ and $T_{\rm yr}$ is the observational
time span in years. The current best upper limit on GW characteristic 
strain is $2.4\times10^{-15}$ \citep{src+13}, which predicts an upper limit on
timing noise of $\sim500$~ns from GW background. Therefore, we
cannot rule out the contribution of gravitational waves in the timing noise.

\subsection{Pulse profile evolution with frequency}
\label{sec:FD}
\begin{figure}
\includegraphics[width=3.4in]{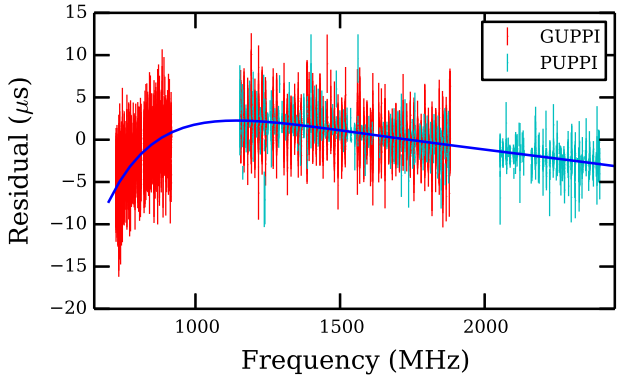} \\ 
\caption {\label{fig:FD} GUPPI and PUPPI post-fit residuals versus frequency when fitted
without the FD model, showing the frequency dependence of the TOAs that is 
not accounted for by the DM. For
clarity, this plot is made using only TOAs with error smaller than 3~$\mu$s.} 
\end{figure}

After removing the dispersion that causes TOA delays proportional to $f^{-2}$,
where $f$ is the observing frequency, we still see small remaining frequency-dependent residuals from wide-band
observations using different instruments and telescopes (Figure \ref{fig:FD}). 
It appears that the low-frequency ($\sim$800~MHz) signals lead the
high-frequency (1400 and 2300~MHz band) signals by microseconds.
The cause of such TOA evolution is not clear. It could be a change in the pulsar's
radiation pattern with frequency, or it could be the use of different
standard profiles in different frequencies. Pulsar radiation of different frequencies may originate from
different parts of the star's magnetosphere, and 
the radiation region of the pulsars' magnetosphere may be slightly distorted,
leading to a frequency-dependent radiation pattern. \citet{pdr14} and \citet{ldc+14} 
extensively discussed this phenomenon and developed TOA extraction techniques
based on phase-frequency 2-D pulse profile matching. This technique is not
yet applied to our data set.

\citet{dfg+13} allowed an arbitrary offset between TOAs taken with different
observing systems and at different frequencies in order to model profile
evolution with frequency.
However, the number of frequency
channels has increased by a factor of ten with the modern wide-band
instruments, making it much harder to mitigate profile-frequency evolution
using frequency channel offsets. 
Instead, we used the FD model, a polynomial of the logarithm of
frequency: $\Delta t_{\rm FD} = \sum_{i=1}^{n} c_i (\log f)^i$
(\citealt{abb+15b}; solid line in Figure
\ref{fig:FD}) to fit for and remove the profile-frequency
evolution, where $\Delta t_{\rm FD}$ is the profile evolution term in unit of
second, $f$ is the observing frequency in unit of GHz and $c_i$ are the
FD model parameters. We employed an F-test with significance value of
0.0027 to determine how many 
FD parameters are needed to model profile frequency evolution. PSR J1713+0747
only requires $n=4$ FD parameters.

\section{Results}
\label{sec:res}

\subsection{Mass measurements}
\label{sec:mass}
\begin{figure*}[!ht]
\centering
\includegraphics[scale=1]{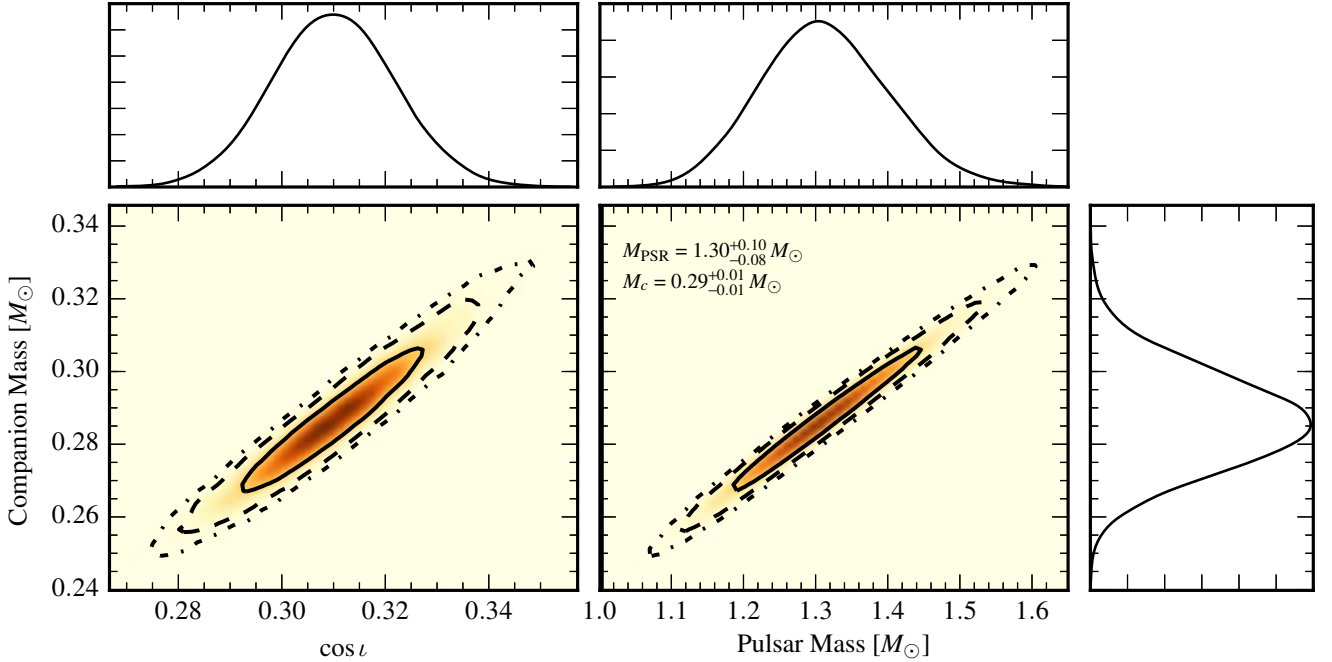} \\ 
\caption {\label{fig:masses}  One and two-dimensional posterior probability
distributions of the cosine of the inclination
angle, pulsar mass, and companion mass from the noise model MCMC including the full nonlinear timing model. The maximum marginalized posterior
value and 1$\sigma$ credible interval is very consistent with the GLS solution
from \textsc{TEMPO}/\textsc{TEMPO2}. The solid, dashed and dashed-dotted lines represent the one, two and three-sigma credible regions, respectively.
} 
\end{figure*}

The timing model of PSR~J1713+0747 has been significantly improved by the 21-year timing effort.
Most notably, the pulsar and the companion masses have been more precisely
constrained (Table \ref{tab:par1} and \ref{tab:par2}, Figure \ref{fig:masses}) through Shapiro delay measurements. The
companion's mass $M_{\rm c} = 0.286\pm0.012$~\Msun\, and the pulsar $M_{\rm
PSR}=1.31\pm0.11$~\Msun\, are in good agreement with the previously measured
values \citep{sns+05}. Furthermore, we have carried out an MCMC run that uses the nonlinear timing
model in order to map out any non-Gaussian correlations in parameter space. We find that the nonlinear
model gives nearly identical results to the GLS method of \textsc{TEMPO}/\textsc{TEMPO2}.
The covariance matrix used in our GLS fitting contains terms come from both the
correlated and the uncorrelated noise; therefore, the timing parameter
uncertainties we get have taken into account the contribution from the noise model.
We note that, without the noise model, the derived pulsar masses would be substantially, and perhaps
unrealistically, lower (with $\Delta M_{\rm PSR}$ ranges from 0.3 to 0.4~$M_{\odot}$) than the
values with the noise model (Table \ref{tab:par1} and \ref{tab:par2}), suggesting that
correlated noise would significantly impact the accuracy of high precision timing analysis.

The pulsar's mass is compatible with the distribution of pulsar masses
in other neutron star-white dwarf systems, and in good
agreement with the distribution of pulsar masses found in recycled binaries
\citep{opns12,kkdt13}. The precise measurement of neutron star masses 
may eventually help us understand the properties of matter of extreme 
density \citep{dpr+10, lat12, afw+13}.

In the standard picture of binary evolution, an MSP with a low-mass white dwarf companion must have been spun up through accretion when the white dwarf was a giant star filling its Roche lobe. 
This should lead to a strong correlation between the binary period and the mass of the white dwarf companion \citep{rpj+95, ts99a, prp02b}. 
Indeed, this picture has been supported the measurements of several pulsar
binary systems \citep[e.g.][]{vbb+01, ktr94, th14, rsa+14}.  
The orbital period and companion mass of PSR~J1713+0747 fit
this correlation very well, thus supporting the standard MSP evolution theory.

\subsection{Intrinsic orbital decay}
\label{sec:obdecay}
We have observed an apparent change in orbital period from PSR~J1713+0747, $\dot{P}_{\rm b} =
(0.36\pm0.17)\times10^{-12}$~s~s$^{-1}$ (Tables \ref{tab:par1} and \ref{tab:par2}).
This orbital period change is not intrinsic to the pulsar binary, but rather the
result of the relative acceleration between the binary and the
observer, i.e.~the combination of differential acceleration in the Galactic gravitational potential
\citep{dt91} and the ``Shklovskii'' effect (\citealt{shk70}) which is caused
by the transverse motion of the pulsar binary relative to Earth. We have good measurements of the distance and proper
motion of the binary system, which allow us to remove these effects and study the system's intrinsic orbital decay.

The apparent change in orbital period due to differential acceleration in
Galactic gravitational potential can be derived from
\begin{equation}
\dot{P}_{\rm b}^{\rm Gal} = \frac{A_{\rm G}}{c} P_{\rm b} =
(-0.10\pm0.02)\times10^{-12}~{\rm s~s^{-1}},
\end{equation}
where $A_{\rm G}$ is the line of sight acceleration of the pulsar binary.
$A_{\rm G}$ is obtained using
Equation 5 in \citet{nt95}, Equation 17 in \citet{lwj+09}, the
local matter density of Galactic disk around solar system \citep{hf04a}
and the Galactic potential model by \citet{rmb+14}.

The Shklovskii effect causes $P_{\rm b}$ to
change by
\begin{equation}
\dot{P}_{\rm b}^{\rm Shk} = (\mu_{\alpha}^2+\mu_{\delta}^2)\frac{d}{c}P_{\rm
b} = (0.65\pm0.02)\times10^{-12}~{\rm s~s^{-1}}.
\end{equation}
Therefore, the pulsar's intrinsic orbital decay is $\dot{P}_{\rm b}^{\rm Int}
= \dot{P}_{\rm b}^{\rm Obs} - \dot{P}_{\rm b}^{\rm Shk} - \dot{P}_{\rm b}^{\rm
Gal} = (-0.20\pm0.17)\times10^{-12}$~s~s$^{-1}$, and is consistent with zero.

Due to the very long $\sim$68 day orbit, the binary's decay due to the
emission of gravitational
radiation is expected to be undetectable: $\dot{P}_{\rm b}^{\rm GR} =
-6\times10^{-18}$~s~s$^{-1}$ \citep{lk05}.  Therefore, the insignificant
intrinsic orbital decay rate is entirely consistent with the
description of quadrupolar gravitational radiation within General
Relativity (GR).

Other than the gravitational radiation, two classical effects could have played a role in
$\dot{P}_{\rm b}^{\rm Int}$. One, $\dot{P}_{\rm b}^{\dot{M}}$, is caused by mass loss in the
binary system, and the other, $\dot{P}_{\rm b}^{\rm T}$, is the contribution
from tidal effects.
The pulsar and the white dwarf both could lose mass due to their magnetic dipole radiation; the maximum
mass loss rate due to this effect can be estimated from the
star's rotational energy loss rate. In the case of the pulsar, $\dot{M}_{\rm
PSR}=\dot{E}/c^2$, measurable through the spin down rate of the pulsar.
The white dwarf generally loses mass at a much lower rate than the pulsar.
Therefore, orbital change due to mass loss can be estimated as $\dot{P}_{\rm
b}^{\dot{M}}\sim 1\times10^{-14}$~s~s$^{-1}$ (\citealt{dt91}; Equation (9) and
(10) of \citealt{fwe+12}). This is an order of magnitude smaller than the measured
uncertainties on $\dot{P}_{\rm b}^{\rm Int}$.
The tidal effect in this binary system is expected to be $\dot{P}_{\rm b}^{\rm
T}\ll3\times10^{-14}$~s~s$^{-1}$ based on the most extreme scenarios (the white
dwarf spins at its break-up velocity and the tidal synchronizing time scale equals the
characteristic age of the pulsar; see Equation 11 in \citealt{fwe+12} and
references therein).
Both of these extra terms are much smaller than the observed uncertainties
on $\dot{P}_{\rm b}^{\rm Int}$.

\subsection{Time Variation of $G$}
\label{sec:Gdot}

Based on the measurement of ``excess'' orbital period change 
$\dot{P}_{\rm b}^{\rm exc}=\dot{P}_{\rm b}^{\rm Int} - \dot{P}_{\rm
b}^{\dot{M}}  - \dot{P}_{\rm b}^{\rm T} - \dot{P}_{\rm b}^{\rm GR}$,
\citet{dgt88} derived a phenomenological limit for $\dot{G}$ without considering the binding energy of the stars: 
$\dot{G}/G\simeq-\dot{P}_{\rm b}^{\rm exc}/(2P_{\rm
b})=(1.0\pm2.3)\times10^{-11}$~yr$^{-1}$ using the timing of binary PSR~B1913+16. 
Since then $\dot{P}_{\rm b}^{\rm exc}$ of pulsar binaries, including 
PSR~J1713+0747, have been used to 
constrain $\dot{G}/G$ \citep{ktr94, nss+05, dvtb08, lwj+09, fwe+12}. 
So far all pulsar observations show $\dot{G}/G$ consistent with being zero, with 
upper limits largely determined by the uncertainties in orbital period change rate, distance, 
and proper motions.
PSR~J1713+0747 has the smallest known $\dot{P}_{\rm b}^{\rm exc}/(2P_{\rm
b})\simeq(-0.5\pm0.9)\times10^{-12}$~yr$^{-1}$ (Section \ref{sec:obdecay}) and is
particularly useful for constraining the time variability of the gravitational
constant.

\begin{figure}
\includegraphics[width=8cm]{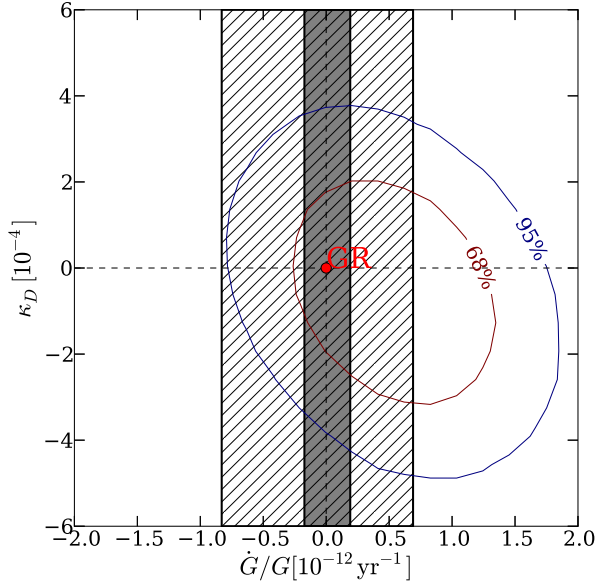} \\ 
\caption {\label{fig:Gdot} Confidence contour of $\dot{G}/G$ and $\kappa_D$
calculated from PSRs~J1012+5307, J1738+0333, and J1713+0747 using an MCMC simulation.
The shaded area marks the 95\% confidence $\dot{G}$ limit from LLR \citep{hmb10}. The gray area
marks the 95\% confidence $\dot{G}$ limit from planetary ephemerides \citep{fle+14}.
} 
\end{figure} 

\citet{nor90}, \citet{lwj+09}, and \citet{fwe+12} showed that a generic test
of $\dot{G}/G$ can be achieved using pulsar
binaries in a more rigorous fashion by incorporating the binding energy of the neutron stars.
The binding energy of a compact star changes with the $G$, resulting in a changing mass, this will also affect the binary orbit.
In a generic form, we could characterize this effect using a self-gravity "sensitivity" parameter $s_p$ \citep{Will93}.
The changing $G$ will now change the orbital period of a pulsar binary system \citep{nor90, lwj+09}:
\begin{equation}
\dot{P}_{\rm b}^{\dot{G}} = -2 \frac{\dot{G}}{G}
\left[1-\left( 1+\frac{m_c}{2M}\right) s_p\right]P_{\rm b}.
\end{equation}
This formalism is more generic in the sense that it incorporates the compactness of the neutron star, 
but it also assumes that non-perturbative effects are absent 
and higher order contributions in the self-gravity sensitivity can be neglected.

Meanwhile, in the framework of an alternative gravitation theory that violates
SEP, a binary system may emit dipole gravitational radiation (\citealt{Will93, Will01, lwj+09, fwe+12} and references
therein). Such effects arise when the two bodies are very different in terms
of their self-gravity, i.e.~their compactness.
Under the aforementioned assumptions of neglecting non-perturbative effects and higher order contributions of self-gravity sensitivity,
this extra dipole radiation could lead to an extra orbital change term:
\begin{equation}
\dot{P}_{\rm b}^{\rm D} \simeq -4\pi\frac{T_{\odot}\mu}{P_{\rm b}}\kappa_D S^2,
\end{equation}
\citep{Will93,lwj+09}, where $T_{\odot}=G{\rm M_{\odot}}/c^3=4.925490947$~${\rm
\mu}$s, $\mu$ is the reduced mass ($m_pm_c/M$) of the system,
$\kappa_D$ is a dipole
gravitational radiation ``coupling constant,'' and $S$ is the difference
between the self-gravity ``sensitivities'' of the two bodies ($S = s_p - s_c$;
$s_p\sim0.1m_p/M_{\odot}$ according to \citealt{de92} ; and $s_c\ll s_p$).
In Einstein's GR $\kappa_D=0$ --- there is no self-gravity induced
dipole gravitational radiation, but it is generally not the case in alternative
theories that violate the SEP.

PSR~J1713+0747 has a wider binary orbit than most other
high-timing-precision pulsar binaries, making its $\dot{P}_{\rm b}^{\rm D}$
very small. Conversely, $\dot{P}_{\rm b}^{\dot{G}}$ is larger when $P_{\rm b}$
is large. This makes PSR~J1713+0747 the best pulsar binary system for constraining
the effect of the changing gravitational constant $\dot{G}$. Limits 
on both $\dot{G}$ and $\kappa_D$ can be estimated in the same fashion as in
\citet{lwj+09}: by solving $\dot{G}$ and $\kappa_D$ simultaneously 
from the equation $\dot{P}_{\rm b}^{\rm exc} = \dot{P}_{\rm b}^{\rm D} +
\dot{P}_{\rm b}^{\dot{G}}$ (Equation 29 of \citealt{lwj+09}) of different
pulsars. We applied this method to four pulsars: PSR J0437$-$4715, PSR J1012+5307, PSR
J1738+0333, and PSR~J1713+0747 using timing parameters reported in
\citet{lwj+09}, \citet{fwe+12}, and this work.
The resulting confidence region of $\dot{G}$ and $\kappa_D$ is shown in Figure
\ref{fig:Gdot}.
We found, at 95\% confidence limit, $\dot{G}/G =
(0.6\pm1.1)\times10^{-12}$~yr$^{-1}$; $\kappa_D=(-0.9\pm3.3)\times10^{-4}$. 
This constraint on $\dot{G}$ is more stringent than
previous pulsar-based constraints \citep{fwe+12},
and close to one of the best constraints of this type
($\dot{G}/G=(-0.07\pm0.76)\times10^{-12}$~yr$^{-1}$) from the Lunar Laser Ranging
(LLR)
experiment \citep{hmb10}, which measured Earth-Moon distance to $\sim10^{-11}$
precision through 39 years of laser ranging.
\citet{fle+14} showed that $\dot{G}/G$ can be constrained to 
$(0.01\pm0.18)\times10^{-12}$~yr$^{-1}$ through the analysis of solar system planetary ephemerides.
The $\kappa_D$ limit, which is not constrained by solar-system tests, is also
slightly improved by using PSR~J1713+0747. 
The pulsar-timing $\dot{G}$ and $\kappa_D$ limits are particularly interesting 
in the testing of SEP-violating alternative theories, because they arise from 
a test using objects of strong self-gravitation. For example, in some classes of the 
scalar-tensor theories, the effect of $\dot{G}/G$ could be significantly enhanced 
in pulsar-white dwarf binaries \citep{wex14}.

\subsection{SEP and PFE}
\label{sec:sep}
The SEP states that the gravitational
effect on a small test body is independent of its constitution, and in
particular, that bodies of different self-gravitation should behave the same in
the same gravitational experiments. This principle is violated in alternative
theories of gravitation like the aforementioned Jordan-Fierz-Brans-Dicke
scalar-tensor theory. The PSR~J1713+0747 binary is an excellent laboratory for testing 
effects of SEP violation. If the SEP is violated, the neutron star and the white
dwarf will be accelerated differently by the Galactic gravitational field, causing
the binary orbit to be polarized toward the center of the Galaxy. The excess 
eccentricity is expected to be (\citealt{ds91}):
\begin{equation}
|\textbf{\textit{e}}_F| = \frac{1}{2}\frac{\Delta\, g_{\bot}
  c^2}{\msF\msG(M_{\rm PSR}+M_{\rm
c})(2\pi/P_{\rm b})^2},
\end{equation}
where $\msF$ is a factor accounts for potential changes in the periastron advance rate
due to deviations from GR, $\msG$ is the effective gravitational constant in the interaction
between the pulsar and the white dwarf, and $g_{\bot}$ is the projection of Galactic acceleration on the orbital plane 
and $\Delta$ is the dimensionless factor that characterizes the significance 
of SEP violation. We assume $\msF\msG\approx G$ here and after.
The Galactic acceleration of the pulsar system is derived from \citet{hf04a, rmb+14}.

GR predicts that there is no preferred reference frame in the universe but this may
be different in alternative theories. 
The Parameterized Post-Newtonian (PPN) formalism parameterized possible
deviations from GR into a set of parameters (see
\citealt{will14} for the list of them), some of which are associated with the
PFE. In this work, we test $\alpha_3$, one of the PFE-related parameters. 
If $\alpha_3 \neq 0$, this would lead to both
the presence of a PFE and the breaking of momentum conservation.
A rotating body would be accelerated perpendicular to its spin
axis and its absolute velocity in the preferred reference frame.
In a pulsar binary, this effect would cause an excess in eccentricity, which can be estimated by (\citealt{de92, bd96}):
\begin{equation}
|\textbf{\textit{e}}_F| = \hat{\alpha}_3 \frac{c_p|\textbf{\textit{w}}|P_{\rm b}^2}{24\pi P}
\frac{c^2}{\msF\msG (M_{PSR}+M_{\rm c})}\sin \beta,
\end{equation}
where $\textbf{\textit{w}}$ is the absolute velocity of the binary system
relative to the preferred frame of reference, typically taken as that of the cosmic microwave background (CMB), $P$ is the pulsar's spin period, $\beta$ is the
angle between $\textbf{\textit{w}}$ and the spin axis of the pulsar, and
$\hat{\alpha}_3$ is the strong-field version of the PPN parameters $\alpha_3$.
Here $\textbf{\textit{w}} = \textbf{\textit{w}}_{\odot} + \textbf{\textit{v}}_{\rm PSR}$, where
$\textbf{\textit{w}}_{\odot}=369\pm0.9$~km~s$^{-1}$ is the velocity of
the solar system relative to the CMB (\citealt{hwh+09}),
and the term $\textbf{\textit{v}}_{\rm PSR}$ is the relative speed of the pulsar to our solar system. 
$\textbf{\textit{v}}_{\rm PSR}$ is only partially known because we can measure
the pulsar system's proper motion on the sky but we cannot measure its
line of sight velocity ($v_{\rm r}$).

Fortunately, many variables in these equations are measurable in the
case of the PSR~J1713+0747 binary. 
It is possible to constrain $\Delta$ and $\hat{\alpha}_3$ using Bayesian techniques 
by assuming certain fiducial priors for $v_{\rm r}$ \citep{sns+05, sfl+05, gsf+11}. 
In our case, we assumed a Gaussian prior for $v_{\rm r}$ centred at zero with
a width equal to the system's proper motion speed.
Based on our 21-year 
timing of J1713+0747 alone, we find 95\% confidence limits on the violations of SEP and
Lorentz invariance $\Delta < 0.01$ and $\hat{\alpha}_3<2\times10^{-20}$, 
slightly improving the single pulsar limits from earlier data on this pulsar
\citep{sns+05, sfl+05, gsf+11}.
Stronger limits can be found by combining the results from
multiple similar pulsar systems \citep{Wex00,sfl+05, gsf+11}.

\section{Summary}
In this paper, we present a comprehensive model of high precision timing observations of
PSR~J1713+0747 that spans 21 years. 
We improved measurements of the pulsar and its companion's masses and the
shape and orientation of the binary orbit. We also detect, for the first time, an apparent
change in orbital period due to Galactic differential accelerations and the Shklovskii effect.
These measurements, when combined with those of other pulsars, 
significantly improve the pulsar timing limit on the rate of change of the gravitational
constant, $\dot{G}$. Although the pulsar constraint is not better than the
best solar system ones, it is nevertheless an independent test using 
extra-solar binary systems thousands of light-years away. The pulsar tests
also could be more constraining for some
classes of alternative gravitational theories that predict stronger non-GR effects in
strong-field regime \citep{wex14}.
The new best pulsar timing limit on $\dot{G}/G$ is 
$(0.6\pm1.1)\times10^{-12}$~yr$^{-1}$ ($<0.033H_0$ based on the 3-$\sigma$ limit), where $H_0$ is the Hubble constant. 
In other words, the change rate of gravitational constant has to be a factor
of at least $31$ (3-$\sigma$ limit) slower than the average expansion rate of
the Universe.

Meanwhile, the precise measurements of PSR J1713+0747's orbital eccentricity and
3D orientation allow us to test the violation of SEP and 
Lorentz invariance with it. We found a single-pulsar 95\% upper limit on 
$\Delta <0.01$, the SEP violation factor, and
$\hat{\alpha}_3<2\times10^{-20}$, the PPN parameter that characterizes
violation of Lorentz invariance. 
Because of the different statistical analysis methods used, our 
$\Delta$ and $\hat{\alpha}_3$ limits  are slightly
different but still consistent with the results of the same tests in previous publications 
\citep{Wex00, sns+05, sfl+05, gsf+11, fkw12}.
Ultimately, the newly discovered pulsar triple system PSR J0337+1715 \citep{rsa+14} 
could yield the best test on SEP violation \citep{fkw12, ssa+15, bbc+15}. In this case 
the inner pulsar-white dwarf binary is orbited by another white dwarf in an
outer orbit, making this system an excellent laboratory for testing
the free fall of a neutron star and white dwarf in an external gravity field.

We studied the time variation of PSR~J1713+0747's DM from 1998 to 2013, and
fitted the structure function of the DM variation with a power law.  
The best-fit power law index is 0.49(5), significantly smaller than the 5/3 
index expected from a ``pure'' Kolmogorov medium. This relatively flat structure
function could be the result of either the lack of long-term DM variations or an
excess of short-term variations. The sudden DM dip around 2008 (Figure
\ref{fig:dmx}) is a good example of such short-term DM variations.
Similar non-Kolmogorov DM variations have been observed in some of the
other NANOGrav pulsars (L. Levin et al. 2015, in preparation.). Evidence of non-Kolmogorov behavior 
in the ISM was also found in the analysis of multi-frequency pulsar scatter times \citep{lkk15}.

As part of our timing modeling, we also included noise contribution
such as jitter and red noise
 using the GLS fit and a covariance matrix that included the
correlated and uncorrelated noise terms.
We found that our timing result is significantly affected by the noise
model, especially the jitter noise, suggesting that the adoption of jitter
modeling may be necessary in other cases of high precision pulsar timing. 
We found that our noise parameters and timing residuals are consistent with the jitter
noise estimates from \citet{sc12} and the timing noise estimate from \citet{sc10}. However,
the scaling law extrapolated from large sample studies of timing noise in \citet{hlk10}
overestimated the timing noise level $\sigma_z({\rm 10 yr})$ in this pulsar.

Our noise model parameters and timing residual RMS (Table \ref{tab:wrms})
provide a crude estimation of the amount of noise in our data. The weighted
root mean square (WRMS; see Table \ref{tab:wrms} for definition) of
the 21-year daily-averaged timing residuals is $\sim 92$~ns. 
Table \ref{tab:wrms} shows a systematic improvement in the timing accuracy of
this pulsar in the last two decades, due to advances in instrumentation.
But the improvements are not as large as expected from the radiometer 
equation, perhaps because of pulse jitter. 

Assuming that the red noise is caused by spin irregularity,
the best-fit spectral index is consistent with
spin irregularity caused by random walks in
either the spin phase or the spin rate of the pulsar, but it does not exclude other explanations
due to its large uncertainty (see bottom panel of Figure \ref{fig:res}).
The observed red noise level is also consistent with the prediction
from the current best upper limit of GW background \citep{src+13}, therefore,
we cannot rule out significant timing noise contribution from the GW background.

\acknowledgements
The authors thank N. Wex, P. C. C. Freire, C. Ng, and J. Cordes for
helpful comments and discussions.
Pulsar research at UBC is supported by an NSERC Discovery Grant and Discovery
Accelerator Supplement, and by the Canadian Institute for Advanced Research.
J. A. E acknowledges support by NASA through Einstein Fellowship grant PF4-150120. 
Some computational work was performed on the Nemo cluster at UWM supported 
by NSF grant No. 0923409. Portions of this research were carried out at the 
Jet Propulsion Laboratory, California Institute of Technology, under a contract 
with the National Aeronautics and Space Administration. 
T. Pennucci is a student at the National Radio Astronomy Observatory.
The NANOGrav project receives support from the National Science Foundation
(NSF) PIRE program award number 0968296. This work was supported by NSF grant 0647820.
The Arecibo Observatory is operated by SRI International under a cooperative
agreement with the National Science Foundation (AST-1100968), and in alliance
with Ana G. M\'{e}ndez-Universidad Metropolitana, and the Universities Space
Research Association.  The National Radio Astronomy Observatory (GBT) is a
facility of the National Science Foundation operated under cooperative
agreement by Associated Universities, Inc.

{\it Facilities:}
\facility{Arecibo Telescope}, \facility{Robert C. Byrd Green Bank Telescope}

\bibliographystyle{apj}
\bibliography{myrefs,journals1,modrefs,psrrefs,crossrefs}

\end{document}